\documentclass[aps,twocolumn,amsmath,amssymb,superscriptaddress,preprintnumbers]{revtex4-1}
\usepackage[pdftex]{graphicx}
\usepackage[outdir=./]{epstopdf}
\usepackage{enumerate,ulem}
\newcommand{\revision}[1]{{#1}}
\begin{document}
\title{Magnetic textures and dynamics in magnetic Weyl semimetals}
\author{Yasufumi Araki}
\affiliation{Advanced Science Research Center, Japan Atomic Energy Agency,
Tokai 319-1195, Japan}
\begin{abstract}
Recent theoretical and experimental attemps have been successful in finding magnetic Weyl semimetal phases,
which show both nodal-point structure in the electronic bands and magnetic orders.
Beyond uniform ferromagnetic or antiferromagnetic orders,
nonuniform magnetic textures, such as domain walls and skyrmions,
may even more enrich the properties of the Weyl electrons in such materials.
This article gives a topical review on interplay between Weyl electrons and magnetic textures in those magnetic Weyl semimetals.
The basics of magnetic textures in non-topological magnetic metals are reviewed first,
and then the effect of magnetic textures in Weyl semimetals is discussed, regarding the recent theoretical and experimental progress therein.
The idea of the fictitious ``axial gauge fields'' is pointed out,
which effectively describes the effect of magnetic textures on the Weyl electrons
and can well account for the properties of the electrons localized around magnetic domain walls.
\end{abstract}

\maketitle

\section{Introduction} \label{sec:introduction}
Magnetism has always been a fundamental concept in materials science.
Along with the development of quantum mechanics,
we have understood the behavior of spins contributing to magnetism,
and have succeeded in designing various magnetic materials that exhibit exotic features useful for applications \cite{Kittel}.
Further technological developments have enabled us to manipulate spins microscopically, called \textit{spintronics} \cite{Zutic_2004,Bader_2010},
which may possibly help us design highly efficient nanoscale devices
that operate at high speed, with low energy consumption, etc.
The discovery of the giant magnetoresistance (GMR) effect
and its aplication to magnetic heads in hard disks are perhaps the most established and successful achievements in spintronics \cite{Baibich_1988,Binasch_1989}.
As well as writing and readout of magnetization in magnetic nanostructures,
detection and manipulation of nanoscale spin textures in magnetic materials have been intensely studied,
aiming to make use of those tiny objects as carriers of information in future devices \cite{Parkin_2008,Fert_2013}.
Various magnetic materials, including both metals and insulators,
have been synthesized and investigated,
to find peculiar features of materials that may be useful for spintronics.

Recent studies have struggled for realizing magnetism in topological materials,
namely the materials whose characteristic electronic structures are
protected by spatial and internal symmetries of the system
and thus classified by topology \cite{Yan_2012,Chiu_2016}.
The most famous class of topological electronic systems
is perhaps \textit{topological insulator},
which shows gapless states on the surface
and is characterized by $\mathbb{Z}_2$ topological invariants defined with the bulk electrons \cite{Hasan_2010,Qi_2011,Bernevig_Book}.
The surface electrons of topological insulators show striking features,
namely the linear (Dirac) dispersion robust under disorder,
strong locking between electron momentum and spin degrees of freedom, etc.,
which have been attracting interest toward their application as well \cite{Pesin_2012}.
By combining these surface Dirac electrons with magnetism,
using magnetic heterostructures or magnetic dopants in experiments \cite{Tokura_2019},
various new phenomena were proposed and successfully observed:
the quantum anomalous Hall effect \cite{Nomura_2011,Yu_2010,Checkelsky_2012,Chang_2013},
universal magneto-optical response \cite{Tse_2010,Okada_2016},
spin-charge conversion \cite{Garate_2010,Shiomi_2014},
current-induced control of magnetization (spin-transfer torque) \cite{Mahfouzi_2012,Mellnik_2014},
etc.

While topological insulators exhibit gapless states on the surface,
recent studies have discovered the classes of materials
showing topologically protected gapless states in the bulk,
which are termed \textit{Dirac} and \textit{Weyl semimetals} \cite{Yan_2017,Burkov_2018,Armitage_2018}.
Dirac/Weyl semimetals,
distinguished by degeneracy of the gapless states,
are characterized by linearly dispersed bands (Dirac/Weyl cones)
around certain band-touching points in momentum space.
These band-touching points, namely the Dirac or Weyl points,
serve as topological objects with monopole charges in momentum space \cite{Volovik_2003},
which give rise to the geometrical phase (Berry phase) of the electrons
and thus contribute to the anomalous transport properties,
such as the anomalous Hall effect or the spin Hall effect \cite{Grushin_2012,Goswami_2013,Burkov_2014,Burkov_2014_2}.
{
Various unusual phenomena,
such as the negative magnetoresistance due to the chiral anomaly \cite{Nielsen_1983,Son_2013,Huang_2015_2,Zhang_2016,Arnold_2016},
quantum oscillations related to the surface-involved Weyl orbits
induced by a magnetic field \cite{Potter_2014,Zhang_2016_2,Moll_2016,Zhang_2017,Zhang_2019},
nonlinear optical responses \cite{Moore_2010,Sodemann_2015,Morimoto_2016,Morimoto_2016_2,Chan_2016,Zhang_2016_3,de_Juan_2017}, etc.,
have been proposed and observed in Weyl semimetals.
Moreover, Weyl semimetals with their Weyl cones tilted to the Fermi surface,
namely \textit{type-II} Weyl semimetals, are also of great interest \cite{Soluyanov_2015}.
Due to their unconventional Fermi surface structure,
modification of magnetic quantum oscillations \cite{OBrien_2016},
superconductivity \cite{Alidoust_2017,Alidoust_2018,Rosenstein_2017,Rosenstein_2018,Shapiro_2018,Hou_2017,Li_2018,Xiao_2017,Das_2018}, optical activity \cite{Mukherjee_2017,Mukherjee_2018,Halterman_2018,Mignuzzi_2019}, etc.,
have been expected and observed in type-II Weyl semimetals.
}

{
Weyl semimetals with broken inversion symmetry
have been experimentally realized in TaAs \cite{Huang_2015,Xu_2015,Lv_2015,Yang_2015}, NbAs \cite{Xu_2015_2}, TaP \cite{Xu_2015_3}, etc.
On the other hand,
}
the Weyl semimetal phase without time-reversal symmetry due to magnetism
has been theoretically proposed from the early days \cite{Burkov_2011,Burkov_2011_2}. 
Over the last few years,
several ferromagnetic and antiferromagnetic materials hosting Weyl electrons
have been numerically demonstrated and experimentally verified \cite{Chen_2014,Kuroda_2017,Liu_2018,Xu_2018,Wang_2018}.
The combination of spin-orbit coupling and magnetism is essential
to retain the band crossing at each Weyl point,
from which we are expecting strong interplay between the magnetism and the electronic properties around the Weyl points in those ``magnetic Weyl semimetals''.

Regarding the above interests,
this article reviews recently developing researches on magnetic Weyl semimetals,
especially focusing on the interplay between nonuniform magnetic textures
and behavior of Weyl electrons.
As seen in ordinary magnetic materials,
magnetic textures, namely spatial or temporal modulation of the (anti)ferromagnetic orders,
enrich the electronic properties in comparison with those under uniform and static orders.
In magnetic Weyl semimetals,
such effects can be efficiently treated with the help of
the idea of fictitious \textit{axial electromagnetic fields} \cite{Liu_2013},
which is the main focus of this review.
The idea of the axial electromagnetic fields, or gauge fields,
was first introduced to describe physics of elementary particles \cite{Peskin_1995},
and is now frequently employed for Dirac and Weyl quasiparticles in condensed matter
to describe effects of various types of spatial and temporal modulations in the systems \cite{Ilan_2019}.

This article is organized as follows.
In Section \ref{sec:spin-texture},
I review common topics about magnetic textures in {conventional} magnetic materials,
and see how they alter the electron transport behavior,
{for later comparison with the case in Weyl semimetals}.
In Section \ref{sec:Weyl-case},
I review current research status about the axial electromagnetic fields in Weyl semimetals
by starting with a minimal model Hamiltonian,
and list up their effects on the structure and transport properties of Weyl electrons.
In Section \ref{sec:domain-wall},
I focus on the properties of magnetic domain walls in magnetic Weyl semimetals,
as typical magnetic textures possibly seen in experiments.
In Section \ref{sec:realistic},
I summarize recent theoretical and experimental achievements
toward realization of magnetic Weyl semimetal phase,
and discuss how the axial electromagnetic field picture can be applied in the proposed systems.
Finally, I conclude this article with some future prospects
about interplay between magnetism and topological electron systems
in Section \ref{sec:conclusion}.

\section{Magnetic textures in normal metals: spin gauge fields \label{sec:spin-texture}}
In this section, I review general theories accounting for
the interplay between magnetic textures and electron dynamics in normal magnetic materials,
{which shall be compared with the treatment in Weyl semimetals for our better understanding}.
Topological magnetic textures
are characterized by topological invariants defined in real space,
which cannot be created or unwound unless they are perturbed by any topological defects.
Thanks to this topological robustness,
these magnetic textures can be macroscopically described as isolated objects;
in the context of spintronics,
a lot of attempts have been made to efficiently manipulate magnetic textures,
in order to make use of them as carriers of information in future devices
such as magnetic racetrack memories \cite{Parkin_2008,Tomasello_2014} and logic gates \cite{Zhang_2015}.

Magnetic domain walls are perhaps the most commonly seen magnetic textures in materials,
since domain structure is necessary to reduce magnetostatic energy from stray field.
Domain wall dynamics has been widely observed in magnetic materials,
which has given us good understanding of the spin torques that can be present in each system \cite{Tatara_2008,Kim_2017}.
Magnetic skyrmions, namely pointlike structures characterized by swirling spin textures inside,
also play an important role in the studies on magnetic materials.
While skyrmions usually form lattice structure, called skyrmion crystal, at ground state,
detection and manipulation of individual skyrmions have been successfully demonstrated
in recent experiments \cite{Nagaosa_2013,Fert_2017,Everschor-Sitte_2018}.

For electric manipulation and detection of such magnetic textures,
we need to understand the interplay between electron transport and magnetic textures.
Electron transport through metallic magnets is affected by magnetic textures 
via the adiabatic phase (Berry phase) accumulated on the electron wave function,
since the electron spin gets eventually modulated by the localized spins in the magnetic textures.
This Berry phase effect can be encoded into the fictitious ``spin electromagnetic fields'' for the electrons,
which act on the majority and minority spin states of the electrons
under spin splitting by the exchange interaction \cite{Korenman_1977,Volovik_1987}.
Topological magnetic textures and their dynamics can host spin electromagnetic fields,
which alter the electron transport through the spin textures,
in a similar manner with the ordinary electromagnetic fields.
Here we briefly review the idea of spin electromagnetic fields
and see some typical phenomena that are twell described by this idea.
(See References ~\cite{Ieda_spincurrent,Tatara_2008,Tatara_2019} for detailed reviews on spin electromagnetic fields.)

Let us start with the minimal model for conduction electrons under magnetic textures,
\begin{align}
  H = \frac{\boldsymbol{p}^2}{2m} + J \boldsymbol{n}(\boldsymbol{r},t)\cdot\boldsymbol{\sigma},
\end{align}
where $m$ is the effective mass of the electron and $\boldsymbol{p}=-i\boldsymbol{\nabla}$ is the momentum operator.
The second term represents the exchange interaction between the electron spin,
denoted by Pauli matrices $\boldsymbol{\sigma}$,
and the local magnetic moment in the magnetic texture,
with its direction given by the unit vector $\boldsymbol{n} = (\sin\theta \cos\phi, \sin\theta\sin\phi, \cos\theta)$,
where $J(>0)$ is the exchange coupling energy.
The key idea to obtain the spin electromagnetic fields
is to rotate the magnetization $\boldsymbol{n}$ to a fixed quantization axis,
namely $z$-axis,
which is achieved by the SU(2) unitary transformation with the matrix
\begin{align}
  U(\boldsymbol{r},t) = e^{i\frac{\theta}{2}\sigma_y} e^{i\frac{\phi}{2}\sigma_z}.
\end{align}
By the unitary transformation
\begin{align}
  H' = U^\dag H U - i U^\dag \partial_t U,
\end{align}
the transformed Hamiltonian $H'$ takes the form
\begin{align}
  H' = \frac{(\boldsymbol{p}+e\pmb{\mathcal{A}})^2}{2m} + J \sigma_z -e\mathcal{A}_0,
\end{align}
where the SU(2) gauge fields $(\pmb{\mathcal{A}},\mathcal{A}_0)$ are given by
\begin{align}
  \pmb{\mathcal{A}} &= -\frac{i}{e} U^\dag \boldsymbol{\nabla} U = -\frac{1}{2e}\left[(\cos\theta \sigma_z - \sin\theta \sigma_x) \boldsymbol{\nabla}\phi + \sigma_y \boldsymbol{\nabla}\theta \right], \\
  \mathcal{A}_0 &= \frac{i}{e} U^\dag \partial_t U = \frac{1}{2e}\left[(\cos\theta \sigma_z - \sin\theta \sigma_x) \dot{\phi} + \sigma_y \dot{\theta} \right].
\end{align}
In general, these gauge fields are in $2 \times 2$ matrix structure.
If the spin splitting $J$ is larger than any other energy scales
so that the interband transition can be neglected,
their diagonal components, namely the projection onto U(1) subspace,
dominantly affect the electron transport.
Focusing on the majority spin state,
which correspond to the lower component of the electron wave function,
the projection of $(\pmb{\mathcal{A}},\mathcal{A}_0)$ onto the majority spin state reads
\begin{align}
  \tilde{\pmb{\mathcal{A}}} = \frac{1}{2e} \cos\theta \boldsymbol{\nabla}\phi , \quad
  \tilde{\mathcal{A}}_0 = -\frac{1}{2e} \cos\theta \dot{\phi}.
\end{align}
These gauge fields yield fictitious electromagnetic fields
\begin{align}
  \tilde{\pmb{\mathcal{E}}} &= -\boldsymbol{\nabla}\tilde{\mathcal{A}}_0 - \partial_t \tilde{\pmb{\mathcal{A}}} = \frac{1}{2e} \sin\theta\left( \dot{\theta}\boldsymbol{\nabla}\phi  - \dot{\phi}\boldsymbol{\nabla}\theta \right), \\
  \tilde{\pmb{\mathcal{B}}} &= \boldsymbol{\nabla} \times \tilde{\pmb{\mathcal{A}}} = -\frac{1}{2e} \sin\theta (\boldsymbol{\nabla} \theta) \times (\boldsymbol{\nabla} \phi), \label{eq:spin-b}
\end{align}
which read in terms of the local magnetization vector $\boldsymbol{n}$,
\begin{align}
  \tilde{\mathcal{E}}_i = \frac{1}{2e} \boldsymbol{n} \cdot (\dot{\boldsymbol{n}} \times \partial_i \boldsymbol{n}), \quad
  \tilde{\mathcal{B}}_i = -\frac{1}{4e} \epsilon_{ijk} \boldsymbol{n} \cdot(\partial_j \boldsymbol{n} \times \partial_k \boldsymbol{n}). \label{eq:spin-b2}
\end{align}
As long as the electron remain on the majority spin state,
which is called the \textit{adiabatic limit},
these spin electromagnetic fields act on the majority-spin electrons
like the ordinary U(1) electromagnetic fields.
Once we incorporate the effect of interband transitions,
we need to consider the nonadiabatic components on the off-diagonal positions of the matrices,
for which I will not go into detail in this article.

From Equation (\ref{eq:spin-b}),
we can see that the spin magnetic field $\tilde{\pmb{\mathcal{B}}}$ is related to the solid angle spanned by the spatial modulation of the magnetization vector $\boldsymbol{n}$.
A typical magnetic texture that yields such a spin magnetic field is a skyrmion \cite{Nagaosa_2013},
since $\boldsymbol{\nabla}\theta$ is in the radial direction and $\boldsymbol{\nabla}\phi$ is in the azimuthal direction in a rotationally symmetric skyrmion.
The spin magnetic field attached to the skyrmions gives rise to
the unconventional Hall transport of the conduction electrons,
which cannot be described as the regular Hall effect related to the applied magnetic field
or the anomalous Hall effect related to the net magnetization of the system \cite{Ye_1999,Tatara_2002}.
This skyrmion-induced Hall effect, called the \textit{topological Hall effect},
was successfully measured in some magnetic materials (e.g. MnSi \cite{Lee_2009,Neubauer_2009}, MnGe \cite{Kanazawa_2011}, etc.)
as a good evidence for the skyrmion crystal phase,
which was originally termed the A-phase.

The spin electric field $\tilde{\pmb{\mathcal{E}}}$ is induced by dynamics of a magnetic texture.
This field exerts a force on the electron,
which is referred to as the \textit{spin motive force},
and drives a conduction current attached to the magnetic texture \cite{Stern_1992,Ryu_1996,Berger_1999,Barnes_2007}.
The spin motive force was experimentally observed
as a current pulse induced by motion of domain walls in magnetic nanowires \cite{Yang_2009,Hai_2009},
and also as a dc voltage induced by ferromagnetic resonance in a comb-patterned ferromagnetic film \cite{Yamane_2011}.

The form of the spin electromagnetic fields shown above applies only to the case
where the electron system in the absence of the exchange interaction is spin-SU(2) symmetric.
In the presence of spin-orbit coupling for the electrons,
the spin electromagnetic fields gets modified from the above form;
under Rashba spin-orbit coupling, for example,
the spin electromagnetic fields acquire additional terms
proportional to the Rashba coupling constant \cite{Kim_2012,Nakabayashi_2014,Tatara_2013}.
The Weyl dispersion can be viewed as
the limit of extermely strong spin-orbit coupling around the band crossing points,
which implies that the effect of spin textures on the Weyl electrons should be
distinct from that in normal metals,
as shall be reviewed in the following sections.

\section{Weyl semimetal and magnetic textures \label{sec:Weyl-case}}
Now we shall see how the interplay between electron transport and magnetic textures
gets altered in magnetic Weyl semimetals.
Due to strong spin-momentum locking around the Weyl points,
the spin gauge field picture introduced in the previous section is
no longer applicable to the Weyl electrons.
Nevertheless, the effect of magnetic textures can still be mapped to fictitious electromagnetic fields for the Weyl electrons,
which are categorized as the \textit{axial} electromagnetic fields,
coupling to the pair of valleys with the opposite signs to each other \cite{Liu_2013,Ilan_2019}.
This picture is available since the location of the Weyl points
depends on the magnetization in magnetic Weyl semimetal,
which is the consequence of strong spin-momentum locking around the Weyl points.
As a result, spatial and temporal modulations of the magnetization
lead to the anomalous responses in the electronic structure and transport \cite{Araki_2018},
which are distinct from those expected in normal metals.

In this section, I review the effect of magnetic textures on the electrons in magnetic Weyl semimetals
based on the idea of the axial magnetic fields.
In the first subsection, I review the general characteristics of Weyl semimetals
with broken time-reversal symmetry,
starting with the toy model.
Although Weyl semimetals are intensely studied and known to show various exotic features,
here I mainly explain the fundamental features of Weyl semimetals
that are necessary for the discussion below on the effect of magnetic textures.
Then I introduce spatial and temporal inhomogeneity in the second subsection,
to see how the fictitious axial electromagnetic fields are defined
associated with magnetic textures.
With the idea of the axial electromagnetic fields,
I summarize in the third subsection
how magnetic textures modulate the electronic structure and transport from macroscopic point of view.
Under the dynamics of magnetic texture, in particular,
we can see pumping of electric charge attached to the magnetic texture,
which is reviewed in the fourth subsection.
Finally, in the last subsection,
I mention the inverse effect,
namely the effect of electron distribution and transport on the magnetic texture dynamics via spin torques,
which can also be described by using the idea of the axial electromagnetic fields.

\subsection{Weyl semimetal with broken time-reversal symmetry} \label{sec:weyl1}

\begin{figure}[tbp]
  \includegraphics[width=\columnwidth]{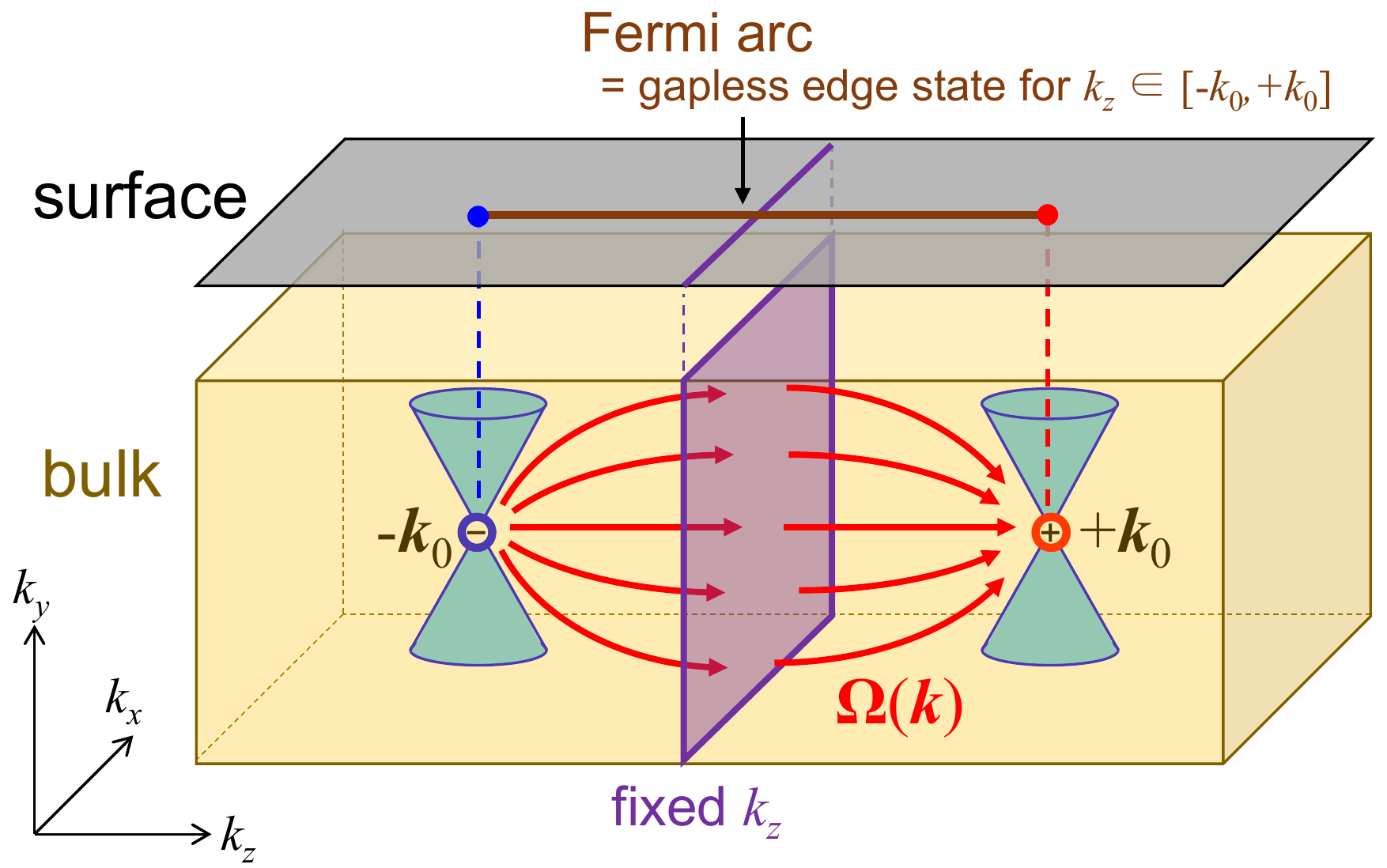}%
  \caption{\label{fig:weyl-nodes}
    Schematic picture of the momentum-space structure of a Weyl semimetal.
    The small circles denote the Weyl points,
    with $+/-$ inside the circles denoting the valley index $\eta$.
    Gapless ``Fermi arc'' modes emerge on the surface Brillouin zone,
    which is the projection of bulk Brillouin zone onto two dimensions.
    See the main text for detail.
    }
\end{figure}

The essence of Weyl semimetal is the linear band touching at certain points,
namely Weyl points, in momentum space.
In contrast to band touching points (Dirac points) in Dirac semimetals,
each of which shows fourfold degeneracy
due to the protection by time-reversal and inversion symmetries,
each Weyl point in Weyl semimetals has twofold degeneracy,
since either time-reversal or inversion symmetry is broken \cite{Burkov_2011_2}.
Each Weyl point behave as a source or sink of Berry curvature in momentum space,
which is defined by $\boldsymbol{\Omega}(\boldsymbol{k}) = i \langle \boldsymbol{\nabla}_{\boldsymbol{k}}u(\boldsymbol{k}) | \times | \boldsymbol{\nabla}_{\boldsymbol{k}}u(\boldsymbol{k}) \rangle$
with the Bloch eigenstate $| u(\boldsymbol{k}) \rangle$,
and thus a topological charge $+1$ or $-1$ is associated with each Weyl point \cite{Volovik_2003,Volovik_2007}.
The net topological charge in momentum space vanishes,
i.e. the numbers of sources and sinks cancel each other over the whole Brillouin zone,
and hence Weyl points with topological charge $+$ and $-$ arise always in pair(s).
This restriction is known as Nielsen--Ninomiya's theorem \cite{Nielsen_1981_1,Nielsen_1981_2},
which has originally been employed to construct lattice models of chiral fermions
for numerical simulations of quantum chromodynamics (QCD) \cite{Chandrasekharan_2004}.

In a Weyl semimetal with inversion symmetry preserved
but time-reversal symmetry broken \cite{Burkov_2011_2,Burkov_2011},
inversion symmetry demands that Weyl points at $\pm \boldsymbol{k}_0$ in momentum space should be paired (see Figure \ref{fig:weyl-nodes}).
Assuming cubic symmetry,
the band dispersion in the vicinity of each Weyl node $\eta \boldsymbol{k}_0 \ (\eta = \pm)$ is described by the minimal Hamiltonian
\begin{align}
  H_\eta(\boldsymbol{k}) = \eta v_{\mathrm{F}} \boldsymbol{\sigma} \cdot (\boldsymbol{k} - \eta \boldsymbol{k}_0), \label{eq:H0}
\end{align}
where the Pauli matrices $\boldsymbol{\sigma}$ denote the electron spin
and $v_\mathrm{F}$ is the Fermi velocity around the Weyl points \cite{Burkov_2011}.
(It should be noted that cubic symmetry is not necessarily present in general,
which implies that $\boldsymbol{\sigma}$ may consist not only of spin degrees of freedom but of other degrees of freedom, such as orbital, sublattice, etc., as well.)
Thus the Weyl-point separation $2\boldsymbol{k}_0$ characterizes
the degree of time-reversal symmetry breaking in the system,
arising from a magnetization, an external magnetic field, etc.
If $\boldsymbol{k}_0 =0$, the Hamiltonian reduces to that of Dirac semimetal
with a Dirac point degenerate at $\boldsymbol{k}=0$.

The electrons residing around the Weyl point at $\eta \boldsymbol{k}_0$,
which we call \textit{valley} $\eta$ for short,
exhibit spin-momentum locking feature.
The electron spin $\boldsymbol{s}_{\eta,\boldsymbol{k}}$ and the momentum $\delta \boldsymbol{k} \equiv \boldsymbol{k} - \eta \boldsymbol{k}_0$
are aligned either parallel or antiparallel,
depending on the valley index $\eta$.
In other words, $\eta$ indicates the helicity (chirality) of the Weyl fermion,
\begin{align}
  \frac{\boldsymbol{s}_{\eta,\boldsymbol{k}} \cdot \delta\boldsymbol{k}}{|\boldsymbol{s}_{\eta,\boldsymbol{k}}| |\delta\boldsymbol{k}|} = \eta
\end{align}
which distinguishes right-handed $(\eta=+)$ and left-handed $(\eta=-)$ modes.
The velocity operator in the valley $\eta$ is given by
\begin{align}
  \boldsymbol{v}_\eta = \frac{\partial H_\eta(\boldsymbol{k})}{\partial \boldsymbol{k}} = \eta v_{\mathrm{F}} \boldsymbol{\sigma}, \label{eq:velocity}
\end{align}
whose expectation value for the Bloch state at $\boldsymbol{k}$
becomes proportional to the spin $\boldsymbol{s}_{\eta,\boldsymbol{k}}$.

The Weyl point $\eta$ possesses a topological charge $-\eta$,
with the Berry curvature
\begin{align}
  \boldsymbol{\Omega}_\eta (\boldsymbol{k}) = -\eta \frac{\delta \boldsymbol{k}}{2|\delta \boldsymbol{k}|^3}
\end{align}
for the electron band (with positive energy).
This Berry curvature distribution leads to the anomalous Hall effect due to the breaking of time-reversal symmetry \cite{Grushin_2012,Goswami_2013,Burkov_2014,Burkov_2014_2}.
Slicing the Brillouin zone by a plane between the two Weyl points,
the first Chern number on this momentum plane,
namely the total Berry flux piercing this plane,
becomes finite, which leads to the anomalous Hall effect in this system.
For instance, if the Weyl points are located
at $\eta \boldsymbol{k}_0 = (0,0,\pm k_0)$ on $k_z$-axis,
the Chern number on the plane at fixed $k_z \in [-k_0,k_0]$ becomes unity,
as shown by the purple plane in Figure \ref{fig:weyl-nodes},
leading to the quantized anomalous Hall conductivity $\sigma^{\mathrm{A}(\mathrm{2D})}(k_z) = e^2/2\pi$ on this plane \cite{Thouless_1982}.
Therefore, summing over the whole Brillouin zone,
the anomalous Hall conductivity in $xy$-plane is given by
\begin{align}
  \sigma^{\mathrm{A}} = \int \frac{dk_z}{2\pi} \sigma^{\mathrm{A}(\mathrm{2D})}(k_z) = \frac{e^2}{4\pi^2} 2k_0,
\end{align}
which is directly related to the Weyl point separation $2\boldsymbol{k}_0$.

The anomalous Hall effect in the bulk is also related with the gapless ``Fermi arc'' states on the surface \cite{Wan_2011,Xu_2011}.
Since the plane at fixed $k_z \in [-k_0,k_0]$ between the two Weyl points
host Chern number 1, as shown above,
there arises a gapless edge mode unidirectionally propagating along the one-dimensional edge of this plane.
The emergence of such a gapless edge state is common in two-dimensional quantum anomalous Hall insulator (Chern insulator) states,
such as the gapped Dirac surface state of a magnetically doped three-dimensional topological insulator \cite{Yu_2010,Checkelsky_2012,Chang_2013,Haldane_1988,Nomura_2011}.
Therefore, on the surface of a Weyl semimetal,
the gapless states emerge for $-k_0 < k_z < k_0$,
connecting the locations of the two Weyl points projected onto the surface.
We can see from this discussion that,
in a Weyl semimetal with broken time-reversal symmetry,
the emergence of the gapless Fermi arc state is
the surface (boundary) counterpart of the anomalous Hall effect in the bulk,
which shall be revisited in Section \ref{sec:domain-wall} to see the effect of magnetic domain walls.
Fermi arcs appear in time-reversal-symmetric Weyl semimetals as well
(although they do not show the anomalous Hall effect),
which have been clearly seen in angle-resolved photoemission spectroscopy (ARPES) measurements \cite{Huang_2015,Xu_2015,Lv_2015,Yang_2015}.
{In Dirac semimetals,
each Dirac point is doubly degenerate and thus serves as a source of two branches of Fermi arcs,
which renders the surface state into closed loops in momentum space,
in contrast to open Fermi arcs in Weyl semimetals
\cite{Liu_2014,Neupane_2014}.}

\subsection{Magnetization and axial electromagnetic fields}
In Weyl semimetals with broken time-reversal symmetry,
the separation of the Weyl points in momentum space
characterizes the effect of the breaking of time-reversal symmetry,
as seen in the previous subsection.
If the breaking of time-reversal symmetry is due to magnetization,
modulation in the magnetization yields shift of the Weyl points in momentum space,
which can be regarded as an effect of a fictitious vector potential.
Based on this idea,
we can consider the effect of magnetic textures and their dynamics
in terms of fictitious electromagnetic fields for the Weyl electrons.

Let us start from the minimal Hamiltonian Equation (\ref{eq:H0}).
If the background magnetization is coupled to the electron spin $\boldsymbol{\sigma}$
by the exchange interaction $J$,
its modulation $\delta \boldsymbol{M}$ gives a perturbation term $J\delta\boldsymbol{M}\cdot\boldsymbol{\sigma}$,
which can be incorporated in the Weyl Hamiltonian as
\begin{align}
  H_\eta(\boldsymbol{k}) = \eta v_{\mathrm{F}}\boldsymbol{\sigma} \cdot \left(\boldsymbol{k} - \eta \boldsymbol{k}_0 - \eta e \boldsymbol{A}_5 \right).
\end{align}
Here $\boldsymbol{A}_5$ is defined by
\begin{align}
  \boldsymbol{A}_5 = -\frac{J}{v_{\mathrm{F}} e} \delta\boldsymbol{M},
\end{align}
which couples to the Weyl electrons in a similar manner to the gauge field,
whereas the sign of its coupling depends on the valley index $\eta$ \cite{Liu_2013}.
In the context of relativistic quantum field theory,
a gauge field coupled to the two chirality channels (right/left-handed) of Dirac fermions
with the signs opposite to each other is termed
the axial gauge field, or the chiral gauge field,
which transformes as an axial vector under the parity operation,
in contrast to the ordinary gauge field as a polar vector \cite{Peskin_1995}.
Since the structure of spin-momentum locking is material-dependent,
the correspondence between magnetization and the axial gauge field
should be modified in some Weyl materials observed in experiments,
which shall be discussed in Section \ref{sec:realistic}.

Just like the ordinary gauge field,
what affects the electronic structure and transport
is not the value of $\boldsymbol{A}_5$ itself but the spatial and temporal structure of $\boldsymbol{A}_5$,
corresponding to the electromagnetic fields.
Let us introduce spatial and temporal structure in $\delta \boldsymbol{M}$,
yielding the axial gauge potential
\begin{align}
  \boldsymbol{A}_5(\boldsymbol{r},t) = -\frac{J}{v_\mathrm{F} e} \delta\boldsymbol{M}(\boldsymbol{r},t).
\end{align}
Here the spatial and temporal variation in $\delta \boldsymbol{M}$ should be moderate enough
to rely on the axial gauge field picture,
since a short-range or high-frequency fluctuation in $\delta \boldsymbol{M}$ may lead to hybridization of the two valleys (Weyl nodes).
Under this condition,
we can define the \textit{axial electric field}
\begin{align}
  \boldsymbol{E}_5(\boldsymbol{r},t) = -\dot{\boldsymbol{A}}_5(\boldsymbol{r},t) = \frac{J}{v_\mathrm{F} e} \dot{\boldsymbol{M}}(\boldsymbol{r},t) \label{eq:E5}
\end{align}
and the \textit{axial magnetic field}
\begin{align}
  \boldsymbol{B}_5(\boldsymbol{r},t) = \boldsymbol{\nabla} \times \boldsymbol{A}_5(\boldsymbol{r},t) = - \frac{J}{v_\mathrm{F} e} \boldsymbol{\nabla} \times  \boldsymbol{M}(\boldsymbol{r},t), \label{eq:B5}
\end{align}
which couple to the two valleys $\eta = \pm$ with the opposite signs $\eta$ \cite{Araki_2018}.
The axial electric field comes from dynamics of the magnetization,
while the axial magnetic field resides at a curled magnetic texture \cite{Liu_2013}.
The simplest example of a magnetic texture that gives rise to the axial magnetic field is a 180-degree domain wall,
as a domain wall always accompanies flip of the magnetization $\boldsymbol{M}$ in real space.
In the presence of these axial electromagnetic fields,
as in the case of normal electromagnetic fields,
the exchange term cannot be absorbed by the local U(1) gauge transformation,
and hence they modulate the electron transport at low energy.

The pseudo-electromagnetic field picture of magnetic textures
is also available for the two-dimensional Dirac electrons on surfaces of topological insulators,
as they show spin-momentum locking around the Dirac point in the surface Brillouin zone \cite{Nomura_2010}.
With this picture, electric charging of magnetic textures,
arising from the fictitious magnetic flux corresponding to the magnetic texture,
was proposed on topological insulator surfaces \cite{Nomura_2010,Wakatsuki_2015}.
It should be noted that there are two major differences.
Since the surface of topological insulator show only a single Dirac cone,
the pseudo-electromagnetic fields from magnetic textures couple to the surface Dirac electrons 
just like the ordinary electromagnetic fields.
Moreover,
only the in-plane two components of magnetization contributes to the pseudo-electromagnetic fields for the surface Dirac electrons,
since the out-of-plane component does not shift but gaps out the surface Dirac point.
These properties are in clear contrast to those in the axial electromagnetic field picture,
which is defined for the pair of Weyl points in three dimensions.

Although the idea of the axial electromagnetic fields introduced here appears similar to the spin electromagnetic fields mentioned in the previous section,
they are different in some aspects.
Conceptually, while the spin electromagnetic fields are obtained by projecting the exchange coupling term to the magnetic textures onto the majority/minority spin states,
the axial electromagnetic fields for the Weyl electrons are derived by projecting them onto the fully spin-momentum-locked states around the Weyl nodes.
The idea of the axial electromagnetic fields is applicable to the limit
of a strong spin-momentum locking and a weak spin splitting,
which is opposite to the situation for the spin electromagnetic fields.
From the phenomenological point of view,
the species of magnetic textures that lead to the axial electromagnetic fields in Weyl semimetals
is much broader than that for the spin electromagnetic fields in normal metals.
{For instance,
the spin magnetic field $\tilde{\pmb{\mathcal{B}}}$ given by Equation (\ref{eq:spin-b2})
in normal magnetic metals requires
at least two-dimensional magnetic textures,
such as skyrmions (see Equation (\ref{eq:spin-b})).
On the other hand, the axial magnetic field $\boldsymbol{B}_5$ can be generated
even from a one-dimensional spin texture,
such as a domain wall
(the case for domain walls shall be discussed in detail in Section \ref{sec:domain-wall}).}
Moreover, dynamics of even a uniform magnetization can lead to the axial electric field $\boldsymbol{E}_5$ in a Weyl semimetal,
which is in a clear contrast with the spin electric field $\tilde{\pmb{\mathcal{E}}}$, namely the spin motive force,
requiring dynamics of a spin texture.
Such a difference arises because the axial gauge potential $\boldsymbol{A}_5$ is tied directly to the local magnetization $\boldsymbol{M}$,
whereas the spin gauge potential comes from the spin connection,
which corresponds to the relative angle between two neighboring spins.
Therefore, the effect of magnetic textures on the electron transport in Weyl semimetals should be
qualitatively different from that in normal magnetic metals,
as long as the electrons are fully spin-momentum-locked on the Fermi surface.

While the idea of the axial electromagnetic fields is introduced to describe the effect of magnetic textures here,
the axial electromagnetic fields can also be reproduced by lattice strain in Dirac and Weyl semimetals \cite{Ilan_2019,Cortijo_2015,Pikulin_2016,Arjona_2018}.
Since a lattice site displacement leads to the modulation of hopping amplitudes and shifts the Dirac/Weyl points,
a lattice strain, namely a spatially nonuniform lattice displacement,
can be regarded as the axial electromagnetic fields in the vicinity of the Dirac/Weyl points.
The effect of the strain-induced axial magnetic field has been intensely studied over recent few years from the theoretical point of view:
Landau quantization \cite{Pikulin_2016},
modulation of the Fermi arc structure \cite{Behrends_2019},
and quantum oscillations due to the Weyl orbits connecting bulk and surface \cite{Liu_2017_2,Pikulin_2018},
has been predicted under the strain-induced axial magnetic field.
It is also proposed that spatial modulation in the chemical composition
of antiperovskite Dirac materials can replicate the pseudomagnetic field
for the Dirac electrons,
since the locations of the Dirac points in those materials are related to
the ratio of the chemical composition \cite{Kariyado_2017}.
The discussions below about the effect of axial gauge fields
can be applied to those systems in almost the similar manner,
while this article will not go into details of them.


\subsection{Charge and current responses to axial electromagnetic fields} \label{sec:responses}
The axial electromagnetic fields couple to the Weyl electrons in the same manner with the realistic electromagnetic fields,
as long as the Weyl nodes can well be treates separately.
Therefore, provided that the length scale of the magnetic texture and the time scale of the magnetization dynamics are much longer than those corresponding to the Weyl-node separation $\boldsymbol{k}_0$,
one may consider the behavior of the electrons within the individual valley $\eta = \pm$,
under the net electromagnetic fields
\begin{align}
  \boldsymbol{E}_\eta = \boldsymbol{E}+\eta\boldsymbol{E}_5, \quad
  \boldsymbol{B}_\eta = \boldsymbol{B}+\eta\boldsymbol{B}_5.
\end{align}
With these electromagnetic fields,
one can estimate the charge and current responses
phenomenologically for each Weyl node \cite{Liu_2013,Araki_2018,Ilan_2019}.
Here I focus on the responses to the fields up to their first order $O(\boldsymbol{E}_\eta,\boldsymbol{B}_\eta)$,
and assume that the Fermi level $\mu$ of the electrons is well defined in equilibrium around the Weyl nodes.
The currents induced by $(\boldsymbol{E},\boldsymbol{B})$ and $(\boldsymbol{E}_5,\boldsymbol{B}_5)$ are summarized in Figure \ref{fig:induced-current} and Table \ref{tab:currents}.

\subsubsection{Equilibrium response}
Let us start from the static system,
without any driving by electric field $\boldsymbol{E}$
or magnetization dynamics characterized by $\boldsymbol{E}_5$.
In the presence of a magnetic field $\boldsymbol{B}_\eta$,
including the axial field $\boldsymbol{B}_5$ from magnetic textures,
it induces the Landau quantization
with the cyclotron frequency $\omega_c = v_\mathrm{F} \sqrt{2e|\boldsymbol{B}_\eta|}$
\cite{Yang_2011},
just like the quantum Hall effect in two-dimensional Dirac electron systems,
such as graphene \cite{CastroNeto_2009,Aoki_book}.
In particular, the zeroth Landau level is linearly dispersed along the direction of $\eta \boldsymbol{B}_\eta$ for each valley $\eta = \pm$,
with the velocity $v_\mathrm{F}$.
The density of states of this zeroth Landau state (per single valley)
is given by $\nu_\eta = e|\boldsymbol{B}_\eta|/4\pi^2 v_\mathrm{F}$,
which is independent of the Fermi energy $\mu$ due to its one-dimensional unidirectional dispersion.
Therefore, if the Fermi level $\mu$ lies below the first Landau level $\epsilon_1(k_z=0) = \omega_c$
so that it may cross only the zeroth Landau levels,
the electric charge is accumulated around the magnetic flux,
with the charge density
\begin{align}
  \rho_B = -e(\nu_+ + \nu_-)\mu = -\frac{e^2}{4\pi^2}\frac{\mu}{v_\mathrm{F}} ( B_+ + B_- )
\end{align}
summed over the two valleys.
(Note that the background charge density without the magnetic field is negligible around the charge neutrality,
since the charge density of free Weyl electrons is proportional to $\mu^3$.)
In particular, even in the absence of the realistic magnetic field $\boldsymbol{B}$,
the axial magnetic flux $\boldsymbol{B}_5(\boldsymbol{r})$ from a magnetic texture leads to the \textit{localized charge},
with its density
\begin{align}
  \rho_B(\boldsymbol{r}) = -\frac{e^2}{2\pi^2} \frac{\mu}{v_\mathrm{F}} |\boldsymbol{B}_5(\boldsymbol{r})| \label{eq:localized-charge}
\end{align}
at the magnetic texture \cite{Pikulin_2016}.
This localized charge is explicitly derived under a one-dimensional domain wall
in terms of the Fermi arc modes localized at the boundary \cite{Araki_2016_2,Araki_2018_2};
see Section \ref{sec:domain-wall} for detail.
Altough the spatially localized charge in metallic regime is inevitably screened once we consider the Coulomb interaction,
the screening effect is smaller than that in normal metals,
since the density of states in topological semimetals becomes small around the band crossing points \cite{Ominato_2018}.

\begin{figure*}[tbp]
  \includegraphics[width=\textwidth]{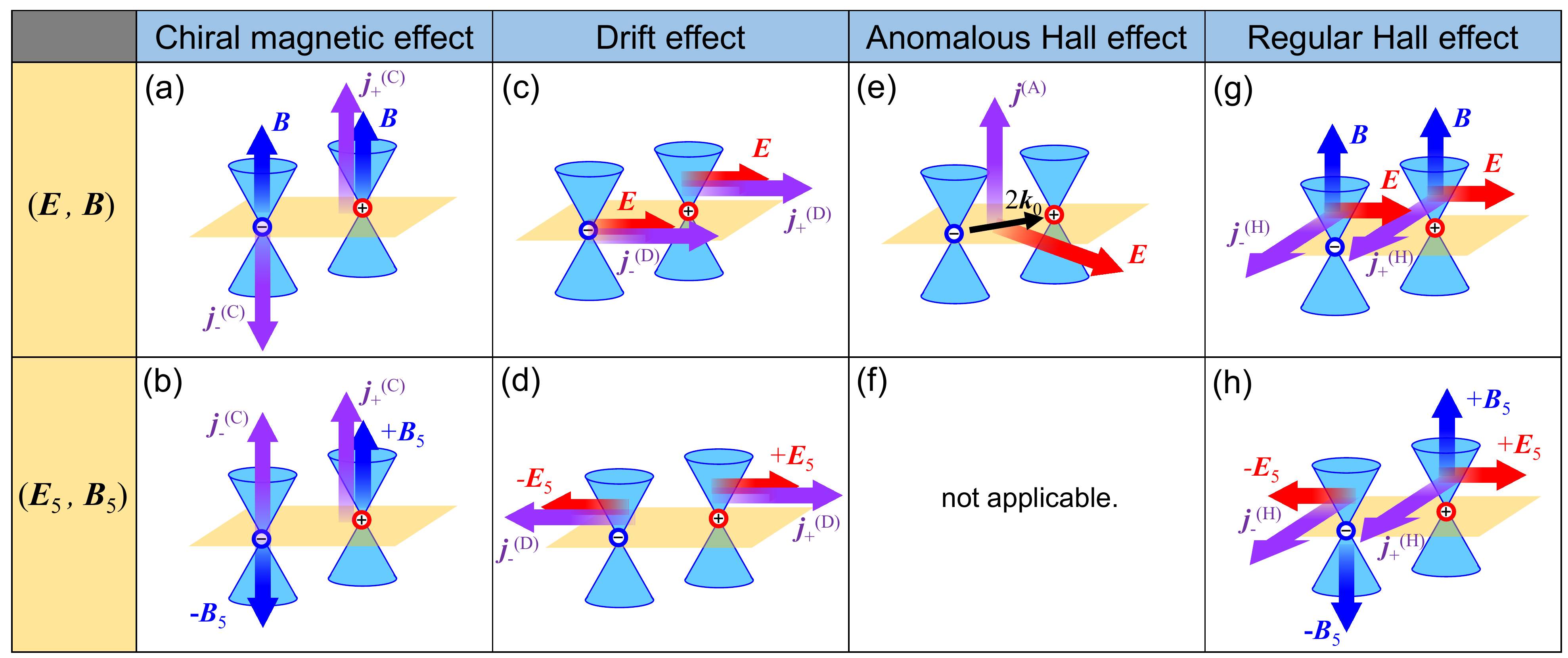}%
  \caption{\label{fig:induced-current} 
  Schematic pictures of the current responses
  induced by the normal electromagnetic fields $(\boldsymbol{E},\boldsymbol{B})$
  and the axial magnetic fields $(\boldsymbol{E}_5,\boldsymbol{B}_5)$.
  Small circles located at the band crossing points denote the valley indices $\eta = \pm$.
  For the cell ``not applicable'', see (ii) in Section \ref{sec:non-equilibrium}.}
\end{figure*}

Since the zeroth Landau level for each valley is dispersed along the direction of $\eta \boldsymbol{B}_\eta$,
the electrons in this Landau level contribute to the current
\begin{align}
  \boldsymbol{j}^{\mathrm{(C)}}_\eta = \rho_{B\eta} v_{\mathrm{F}} \eta \hat{\boldsymbol{B}}_\eta = -\frac{e^2}{4\pi^2} \mu \eta \boldsymbol{B}_\eta
\end{align}
for each valley $\eta$ \cite{Grushin_2016,Pikulin_2016},
where a bold symbol with a ``hat'' denotes {its} unit vector ($\hat{\boldsymbol{X}} \equiv \boldsymbol{X}/|\boldsymbol{X}|$).
Therefore, the net current induced by the magnetic fields reads
\begin{align}
  \boldsymbol{j}^{\mathrm{(C)}} = -\frac{e^2}{4\pi^2} \mu \left(\boldsymbol{B}_+ - \boldsymbol{B}_- \right) = -\frac{e^2}{2\pi^2} \mu \boldsymbol{B}_5, \label{eq:j-chiral}
\end{align}
which depends only on the axial magnetic field $\boldsymbol{B}_5$ (see Figure \ref{fig:induced-current}(b)).
This effect is named
the \textit{chiral pseudomagnetic effect} or the \textit{chiral axial magnetic effect} in literatures \cite{Araki_2018,Zhou_2012,Huang_2017}.

The chiral pseudomagnetic effect is the axial counterpart of the chiral magnetic effect,
namely the current generation by a magnetic field $\boldsymbol{B}$
in the presence of chemical potential imbalance between two valleys (left/right-handed fermions),
which has long been known in the context of chiral fermions at high energy
(heavy ion collisions, neutron stars, etc.) \cite{Vilenkin_1980,Fukushima_2008,Kharzeev_2008,Kharzeev_2014,Son_2012}.
It is shown that the chiral magnetic effect in equilibrium is absent in lattice systems,
since the imbalance in the Fermi levels of two valleys within the same lattice model is not available in equilibrium \cite{Vazifeh_2013,Yamamoto_2015}.
The magnetic field $\boldsymbol{B}$ in equilibrium contributes
only to the axial current (see Figure \ref{fig:induced-current}(a)):
$\boldsymbol{j}_5^{\mathrm{(C)}} = \boldsymbol{j}_+^{\mathrm{(C)}} - \boldsymbol{j}_-^{\mathrm{(C)}} = -(e^2/2\pi^2) \mu \boldsymbol{B}$ \cite{Taguchi_2015}.
The chiral magnetic effect is thus sought for in inequilibrium \cite{Chen_2013,Ma_2015,Sekine_2016,Zhong_2016};
the negative magnetoresistance in Dirac/Weyl semimetals is a typical inequilibrium phenomenon
that stems from the chiral magnetic effect \cite{Nielsen_1983,Son_2013}.

The chiral pseudomagnetic effect due to the axial magnetic field $\boldsymbol{B}_5$,
on the other hand,
can locally induce an equilibrium current.
One can qualitatively understand this local equilibrium current
in connection with the orbital magnetization $\boldsymbol{M}_{\mathrm{orb}}$ of the electron system,
by the relation $\boldsymbol{j}(\boldsymbol{r}) = \boldsymbol{\nabla} \times \boldsymbol{M}_{\mathrm{orb}}(\boldsymbol{r})$ \cite{Grushin_2016}.
\revision{
The logic is threefold:
(i) The orbital magnetization $\boldsymbol{M}_{\mathrm{orb}}$ of a Weyl semimetal
appears proportional to the spin magnetization $\boldsymbol{M}$.
(This can be easily understood if $\boldsymbol{M}$ is uniform,
as the surfaces host a circulating current carried by the Fermi-arc states.)
(ii) If there is a spatial inhomogeneity in $\boldsymbol{M}(\boldsymbol{r})$,
the orbital magnetization $\boldsymbol{M}_{\mathrm{orb}}(\boldsymbol{r})$ is also inhomogeneous,
leading to the local current $\boldsymbol{j}(\boldsymbol{r}) = \boldsymbol{\nabla} \times \boldsymbol{M}_{\mathrm{orb}}(\boldsymbol{r})$ present in the bulk.
(iii) Since $\boldsymbol{\nabla} \times \boldsymbol{M}(\boldsymbol{r})$ corresponds to the axial magnetic field $\boldsymbol{B}_5(\boldsymbol{r})$,
we can regard this local current $\boldsymbol{j}(\boldsymbol{r})$ as the bound current localized at the axial magnetic flux $\boldsymbol{B}_5(\boldsymbol{r})$.
}
(This bound current was calculated explicitly under a magnetic domain wall \cite{Araki_2016_2}.)
It was also numerically demonstrated on lattice models
that an axial magnetic field corresponding to lattice strain leads to an equilibrium current localized at the torsion axis,
although the net current over the whole system is zero \cite{Pikulin_2016,Gorbar_2017}.


\begin{table}[tbp]
    \caption{Summary of the current responses generated by the normal electromagnetic fields (EMFs) $(\boldsymbol{E},\boldsymbol{B})$ and the axial EMFs $(\boldsymbol{E}_5,\boldsymbol{B}_5)$. For the cell ``not applicable'', see (ii) in Section \ref{sec:non-equilibrium}. \label{tab:currents}}%
    \begin{tabular}{ccc}
    Classification & Normal EMFs & Axial EMFs \\
    Chiral magnetic effect & $\boldsymbol{j}_5^{\mathrm{(C)}} \propto \boldsymbol{B}$ & $\boldsymbol{j}^{\mathrm{(C)}} \propto \boldsymbol{B}_5$ \\
    Drift effect & $\boldsymbol{j}^{\mathrm{(D)}} \propto \boldsymbol{E}$ & $\boldsymbol{j}_5^{\mathrm{(D)}} \propto \boldsymbol{E}_5$ \\
    Anomalous Hall effect & $\boldsymbol{j}^{\mathrm{(A)}} \propto \hat{\boldsymbol{k}}_0 \times \boldsymbol{E}$ & not applicable \\
    Regular Hall effect & $\boldsymbol{j}^{\mathrm{(H)}} \propto \hat{\boldsymbol{B}} \times \boldsymbol{E}$ & $\boldsymbol{j}^{\mathrm{(H)}} \propto \hat{\boldsymbol{B}}_5 \times \boldsymbol{E}_5$
  \end{tabular}
\end{table}

\subsubsection{Nonequilibrium response} \label{sec:non-equilibrium}

When the electric fields $\boldsymbol{E}_\eta$ are switched on,
the electron distribution is driven to nonequilibrium state
and it gives rise to various current responses.
$\boldsymbol{E}_\eta$ consists of the normal electric field $\boldsymbol{E}$ and the axial electric field $\boldsymbol{E}_5$ corresponding to the dynamics of magnetization.
Up to the linear response to $\boldsymbol{E}_\eta$,
the current response in Weyl semimetals is classified into three contributions:
(i) the drift current, (ii) the anomalous Hall current, and (iii) the regular Hall current \cite{Araki_2018}.
I summarize these contributions below,
first using the generalized electric field $\boldsymbol{E}_\eta$,
and then limiting it to the axial electric field $\boldsymbol{E}_5$
corresponding to the magnetization dynamics in the magnetic Weyl semimetal.

(i) The \textit{drift effect} is the current response longitudinal to the applied electric field $\boldsymbol{E}_\eta$,
given by
\begin{align}
  \boldsymbol{j}_\eta^{\mathrm{(D)}} = \sigma_\eta^{\mathrm{D}} \boldsymbol{E}_\eta
\end{align}
for each valley $\eta$ (see Figure \ref{fig:induced-current}(a)).
Here $\sigma_\eta^{\mathrm{D}}$ denotes the longitudinal conductivity for valley $\eta$,
which is well defined as long as the Fermi surfaces of the two valleys are well separated in momentum space
so that the intervalley scattering can be negligible.
Under this condition, $\sigma_\eta^{\mathrm{D}}$ can be estimated semiclassically:
$\sigma_\eta^{\mathrm{D}} = e^2 v_{\mathrm{F}}^2 D(\mu) \tau/3$ for spherically symmetric Weyl dispersion,
where $D(\mu) = \mu^2/2\pi^2 v_{\mathrm{F}}^3$ is the density of states (per single Weyl cone)
and $\tau$ is the transport relaxation time.

In the absence of the realistic electric field $\boldsymbol{E}$,
the current and the axial current induced by the axial electric field $\boldsymbol{E}_5$
are given by
\begin{align}
  \boldsymbol{j}^{\mathrm{(D)}} = \left( \sigma_+^{\mathrm{D}} - \sigma_-^{\mathrm{D}}\right) \boldsymbol{E}_5, \quad
  \boldsymbol{j}_5^{\mathrm{(D)}} = \left( \sigma_+^{\mathrm{D}} + \sigma_-^{\mathrm{D}}\right) \boldsymbol{E}_5
\end{align}
In magnetic Weyl semimetals,
the two valleys have the identical structure due to its inversion symmetry.
Therefore, the longitudinal conductivities for the two valleys are equal,
and hence the axial electric field induces no net current $\boldsymbol{j}^{\mathrm{(D)}} =0$ (see Figure \ref{fig:induced-current}(d));
it only drives the axial current $\boldsymbol{j}_5^{\mathrm{(D)}}$,
which corresponds to the spin accumulation $\langle\boldsymbol{\sigma} \rangle$ if spin and momentum are fully locked.

(ii) The \textit{anomalous Hall effect} is the current response
transverse to the applied electric field.
In contrast to the regular Hall effect,
the anomalous Hall effect does not require any magnetic field
and arises from other time-reversal symmetry breaking effect,
such as magnetization \cite{Karplus_1954}.
As is well known, the anomalous Hall effect can be classified into
the extrinsic effect coming from the asymmetric scattering by impurities
and the intrinsic effect due to the anomalous velocity driven by the nonzero Berry curvature in momentum space.
In magnetic Weyl semimetals, as shown in Section \ref{sec:weyl1},
the intrinsic anomalous Hall effect arises from the separation of two Weyl points:
the induced current is given by
\begin{align}
  \boldsymbol{j}^{\mathrm{(A)}} = \sigma^{\mathrm{A}} \hat{\boldsymbol{k}}_0 \times \boldsymbol{E},
\end{align}
where the anomalous Hall conductivity is tied to the Weyl-point separation $2 \boldsymbol{k}_0$ as $\sigma^{\mathrm{A}} = (e^2/2\pi^2) |\boldsymbol{k}_0|$
(see Figure \ref{fig:induced-current}(e)).
Since this intrinsic Hall effect comes from all the occupied states below the Fermi level,
it cannot be separated to the individual valleys.
Thus the ``axial electric field'' picture cannot be applied to see the intrinsic anomalous Hall effect contribution (the cells ``not appliable'' in Table \ref{tab:currents} and Figure \ref{fig:induced-current}(f));
it may depend on the modulation of band structure by the magnetic texture dynamics away from the Weyl points.

(iii) The \textit{regular Hall effect} is the current response induced by the electric field $\boldsymbol{E}_\eta$ in the presence of the magnetic field $\boldsymbol{B}_\eta$:
provided the regular Hall conductivity $\sigma^{\mathrm{H}}_\eta$ is defined for the  valleys $\eta$ separately,
the current is written as
\begin{align}
  \boldsymbol{j}^{\mathrm{(H)}}_\eta = \sigma^{\mathrm{H}}_\eta \hat{\boldsymbol{B}}_\eta \times \boldsymbol{E}_\eta.
\end{align}
The Hall conductivity $\sigma^{\mathrm{H}}_\eta$ can be estimated
in either semiclassical or quantum regime,
depending on the magnetic field strength $|\boldsymbol{B}_\eta|$,
the Fermi energy $\mu$, and the level broadening effect $\tau^{-1}$ by impurities.
\begin{itemize}
  \item \textit{Quantum regime}:
  If the Fermi level $\mu$ and the level broadening $\tau^{-1}$ are both below the first Landau level $\epsilon_1(k_z=0) = \omega_c$,
  only the zeroth Landau level contributs to the regular Hall effect.
  The Hall conductivity in this regime is $\sigma^{\mathrm{H(q)}}_\eta = (e^2/4\pi^2) (\mu/v_{\mathrm{F}})$,
  which is the three-dimensional counterpart of the two-dimensional quantum Hall conductivity $e^2/2\pi$ \cite{Yang_2011}.
  \item \textit{Semiclassical regime}:
  If the Fermi level $\mu$ is far beyond the Landau-level spacing $\sim \omega_{\mathrm{c}}$,
  we can treat the Hall transport unquantized.
  The semiclassical Hall conductivity from the Boltzmann transport theory
  is given by $\sigma^{\mathrm{H(c)}}_\eta = -(\tau^2 e^3 \mu/6\pi^2) |\boldsymbol{B}_\eta|$.
\end{itemize}

In the absence of the realistic electromagnetic fields,
the axial electromagnetic fields $(\boldsymbol{E}_5,\boldsymbol{B}_5)$,
corresponding to the dynamics and texture of the magnetization in a magnetic Weyl semimetal,
are the only source of the regular Hall effect.
Since both $\boldsymbol{E}_5$ and $\boldsymbol{B}_5$ couple to the valley $\eta$ by the sign $\eta$,
the regular Hall current is induced in the same direction for the two valleys (see Figure \ref{fig:induced-current} (h)),
\begin{align}
  \boldsymbol{j}^{\mathrm{(H)}}_\eta = \sigma^{\mathrm{H}}_\eta \eta\hat{\boldsymbol{B}}_5 \times \eta\boldsymbol{E}_5 = \sigma^{\mathrm{H}}_\eta \hat{\boldsymbol{B}}_5 \times \boldsymbol{E}_5.
\end{align}
In particular, if the axial magnetic field $\boldsymbol{B}_5$ is strong enough to reach the quantum regime,
the net regular Hall current for the two valleys
is simply given as
\begin{align}
  \boldsymbol{j}^{\mathrm{(H)}} = \left(\sigma^{\mathrm{H}}_+ + \sigma^{\mathrm{H}}_- \right) \hat{\boldsymbol{B}}_5 \times \boldsymbol{E}_5 =\frac{e^2}{2\pi^2} \frac{\mu}{v_{\mathrm{F}}} \hat{\boldsymbol{B}}_5 \times \boldsymbol{E}_5, \label{eq:j-Hall-quantum}
\end{align}
which is independent of the field strength $|\boldsymbol{B}_5|$.

In addition to the current responses mentioned above,
the Weyl fermions are subject to the \textit{chiral anomaly},
namely the violation of charge conservation in the presence of electromagnetic fields.
The idea of chiral anomaly was originally established
in the context of relativistic field theory,
to account for the anomalous decay of a pion \cite{Adler_1969,Bell_1969},
and recently it has been intensely applied to Dirac and Weyl electrons in materials \cite{Burkov_2015}.
If we naively focus on the electron dynamics around the Fermi surface,
the chiral anomaly states that charge conservation within each valley (chirality) $\eta = \pm$ is violated by the electromagnetic fields $(\boldsymbol{E}_\eta,\boldsymbol{B}_\eta)$ as
\begin{align}
  \partial_t \rho_\eta + \boldsymbol{\nabla} \cdot \boldsymbol{j}_\eta = -\frac{e^3}{4\pi^2} \boldsymbol{E}_\eta \cdot \boldsymbol{B}_\eta,
\end{align}
which is called the covariant anomaly \cite{Landsteiner_2014,Landsteiner_2016}.
The covariant anomaly, however, appears to violate the conservation of net charge as
\begin{align}
  \partial_t \rho + \boldsymbol{\nabla} \cdot \boldsymbol{j} = -\frac{e^3}{2\pi^2} (\boldsymbol{E} \cdot \boldsymbol{B}_5 + \boldsymbol{E}_5 \cdot \boldsymbol{B}).
\end{align}
This unphysical situation is resolved by supplementing the regularization-dependent terms in the Lagrangian,
namely the ``Bardeen polynomials'' \cite{Bardeen_1969,Alvarez-Gaume_1984}.
This treatment yields the charge-conserving relation
\begin{align}
  \partial_t \rho + \boldsymbol{\nabla} \cdot \boldsymbol{j} = 0, \quad
  \partial_t \rho_5 + \boldsymbol{\nabla} \cdot \boldsymbol{j}_5 = -\frac{e^3}{2\pi^2} \left[\boldsymbol{E} \cdot \boldsymbol{B} + \frac{1}{3}\boldsymbol{E}_5 \cdot \boldsymbol{B}_5\right],
\end{align}
which is called the consistent anomaly.
The Bardeen polynomials correspond to the current and charge
carried by the occupied states away from the valleys in the context of Weyl semimetals on lattice,
including the anomalous Hall current $\boldsymbol{j}^{\mathrm{(A)}}$ mentioned above \cite{Gorbar_2017,Behrends_2019}.
The consistent anomaly leads to the chiral charge imbalance,
namely the imbalance in the numbers of right-handed and left-handed fermions,
which gives rise to several observable phenomena;
for instance, the negative magnetoresistance in Dirac/Weyl semimetals
is described as the combined effect of the chiral charge imbalance from the anomaly
and the current induction by the chiral magnetic effect \cite{Son_2013}.
While the chiral anomaly is present under the axial electromagnetic fields as well,
I will not go into detail in the discussions below,
since it does not modulate the net charge and current profiles at linear response to those fields.

\subsection{Charge pumping by magnetic texture dynamics}
Based on the forementioned list of current responses to the axial electromagnetic fields,
we are now ready to discuss the charge and current responses to the dynamics of magnetic textures in a Weyl semimetal.
If there is a magnetic texture $\boldsymbol{M}(\boldsymbol{r},t)$
that is spatially localized and temporally modulating,
we can define the axial electric field $\boldsymbol{E}_5$ and the axial magnetic field $\boldsymbol{B}_5$
localized at the magnetic texture [Equations(\ref{eq:E5}) and (\ref{eq:B5})].
If the spatial and temporal modulations of the magnetic texture are moderate enough
compared to the mean-free path and time of the electrons,
we can consider the axial electromagnetic fields \textit{locally uniform}
within these scales,
which enables us to use the macroscopic response picture summarized above
in the vicinity of the magnetic texture.
Here I focus on the local current distribution
in response to the axial electromagnetic fields,
and derive the dynamics of electric charge distribution
accompanied with the magnetic texture dynamics \cite{Araki_2018,Kurebayashi_2019}.

In the absence of real electromagnetic fields $(\boldsymbol{E},\boldsymbol{B})$,
the chiral pseudomagnetic effect $\boldsymbol{j}^{\mathrm{(C)}}$ and the regular Hall effect $\boldsymbol{j}^{\mathrm{(H)}}$
are the only contribution to the net charge current $\boldsymbol{j}_{\mathrm{ind}}(\boldsymbol{r},t)$,
up to linear response to the axial electromagnetic fields $(\boldsymbol{E}_5,\boldsymbol{B}_5)$.
Since the chiral anomaly from $\boldsymbol{E}_5 \cdot \boldsymbol{B}_5$
leads only to the chiral charge imbalance
and does not violate the net charge conservation,
we can use the charge conservation relation
\begin{align}
  \partial_t \delta \rho(\boldsymbol{r},t) = -\boldsymbol{\nabla} \cdot \boldsymbol{j}_{\mathrm{ind}}(\boldsymbol{r},t) = -\boldsymbol{\nabla} \cdot \left[ \boldsymbol{j}^{\mathrm{(C)}}(\boldsymbol{r},t) + \boldsymbol{j}^{\mathrm{(H)}}(\boldsymbol{r},t) \right] \label{eq:charge-conservation}
\end{align}
to estimate the charge density profile $\delta \rho(\boldsymbol{r},t)$
modulated (pumped) by the magnetic texture dynamics.
Since the chiral pseudomagnetic effect contribution
$\boldsymbol{j}^{\mathrm{(C)}} \propto \boldsymbol{B}_5 \propto \boldsymbol{\nabla} \times \boldsymbol{A}_5$
is divergence-free,
the only contribution to the pumped charge $\delta \rho$
is from the regular Hall current $\boldsymbol{j}^{\mathrm{(H)}}$.
If the magnetic texture is well localized
so that the axial magnetic field should be strong enough,
only the zeroth Landau level contributes to the charge pumping;
substituting $\boldsymbol{j}^{\mathrm{(H)}}$ in the quantum regime [Equation (\ref{eq:j-Hall-quantum})]
to the charge conservation relation Equation (\ref{eq:charge-conservation}),
one obtains the relation
\begin{align}
  \partial_t \delta\rho(\boldsymbol{r},t) = \frac{e^2}{2\pi^2} \frac{\mu}{v_{\mathrm{F}}} \left[ \hat{\boldsymbol{B}}_5 \cdot (\boldsymbol{\nabla} \times \boldsymbol{E}_5) - \boldsymbol{E}_5 \cdot (\boldsymbol{\nabla} \times \hat{\boldsymbol{B}}_5)\right], \label{eq:charge-conservation-2}
\end{align}
from which one can derive the time evolution of the charge distribution $\delta \rho(\boldsymbol{r},t)$
induced by the magnetic texture dynamics.

Assuming there is no curl in the direction of the axial magnetic field $\hat{\boldsymbol{B}}_5$,
the second term in the right hand side of Equation (\ref{eq:charge-conservation-2}) vanishes
and this relation can be further simplified.
While it is difficult to rewrite this condition for the magnetic texture $\boldsymbol{M}(\boldsymbol{r})$ in general,
we can consider some extreme cases that satisfy this condition:
if $\boldsymbol{M}(\boldsymbol{r})$ is aligned within a certain plane,
which can be realized under a strong easy-plane magnetic anisotropy,
$\boldsymbol{B}_5$ points perpendicular to this plane
and thus $\hat{\boldsymbol{B}}_5$ becomes homogeneous.
In such cases,
by using the general relation $\boldsymbol{\nabla} \times \boldsymbol{E}_5 = -\partial_t \boldsymbol{B}_5$,
we obtain a further simplified relation
\begin{align}
  \delta \rho(\boldsymbol{r},t) = -\frac{e^2}{2\pi^2} \frac{\mu}{v_{\mathrm{F}}} |\boldsymbol{B}_5(\boldsymbol{r},t)| + \mathrm{const.}
\end{align}
This relation is consistent with Equation (\ref{eq:localized-charge}) obtained in equilibrium.
Therefore, we can see that the localized charge $\rho_B$
arising from the axial magnetic flux $\boldsymbol{B}_5$
moves together with the dynamics of magnetic texture.

The charge pumping effect in magnetic Weyl semimetal appears
similar to the current induction by the spin motive force in normal magnetic metals,
mentioned in Section \ref{sec:spin-texture}.
Their difference can be understood
by considering the work (energy transfer) on the electrons exerted by the magnetic texture.
The spin motive force in normal metals
act on an electron as the drift force $-e \tilde{\pmb{\mathcal{E}}}$ by the spin electric field $\tilde{\pmb{\mathcal{E}}}$.
Since the drift force exerts a work on the transported electron,
the energy of the magnetic texture dynamics is eventually transferred to the ensemble of electrons,
which is usually dissipated via the electron scattering by impurities.
On the other hand, in magnetic Weyl semimetals,
the driving force from the axial electromagnetic fields $(\boldsymbol{E}_5,\boldsymbol{B}_5)$
is the Lorentz force $-e \dot{\boldsymbol{r}} \times \boldsymbol{B}_5$,
which is perpendicular to the path of the electron.
Therefore, it does not exert a work on the electrons,
and the magnetic texture dynamics does not lose its energy by this pumping effect.
The energy is dissipated only via the Gilert damping of the constituent spins in the magnetic texture,
so that the magnetic texture dynamics does not significantly heat up the Weyl electrons.
In this sense, this pumping effect in magnetic Weyl semimetal
can be regarded ``adiabatic'' \cite{Araki_2018}.

\subsection{Field-induced dynamics of magnetic textures}\label{sec:torque}
So far we have seen that the electron dynamics driven by magnetic texture dynamics
can be understood with the idea of the axial electromagnetic fields.
Similarly, driving of magnetic texture dynamics by the electrons,
namely the spin transfer torque and the spin-orbit torque,
can also be formulated by using the idea of the axial electromagnetic fields \cite{Kurebayashi_2019}.

Generally speaking, the spin torque is induced
by the electron spin accumulation $\langle \boldsymbol{\sigma} \rangle$.
It gives an effective magnetic field $J \langle \boldsymbol{\sigma} \rangle$
on the magnetization vector $\boldsymbol{n}$ via the exchange interaction $J$,
leading to the torque $\boldsymbol{T} = J \langle \boldsymbol{\sigma} \rangle \times \boldsymbol{n}$.
In a Weyl semimetal of the toy model Equation (\ref{eq:H0}), in particular,
the spin accumulation is given equivalent to the axial current,
\begin{align}
  \boldsymbol{j}_5 = -e \sum_{\eta=\pm} \eta \langle \boldsymbol{v}_\eta \rangle_\eta = -e v_{\mathrm{F}} \sum_{\eta=\pm} \langle \boldsymbol{\sigma} \rangle_\eta = -e v_{\mathrm{F}} \langle \boldsymbol{\sigma} \rangle
\end{align}
by using Equation (\ref{eq:velocity}),
where $\langle \cdot \rangle_\eta$ denotes the expectation value within valley $\eta$.
Therefore, we here need to focus on the axial current response
to estimate the field-induced torques on the magnetic texture.

When an electric field $\boldsymbol{E}$ is applied to the magnetic texture,
there arises a spin-transfer torque described by the regular Hall effect \cite{Kurebayashi_2019}:
since the magnetic texture accompanies the axial magnetic field $\boldsymbol{B}_5$,
the current driven by the regular Hall effect is an axial current,
\begin{align}
  \boldsymbol{j}_5^{\mathrm{(H)}} = \boldsymbol{j}_+^{\mathrm{(H)}} - \boldsymbol{j}_-^{\mathrm{(H)}} = \frac{e^2}{2\pi^2} \frac{\mu}{v_{\mathrm{F}}} \hat{\boldsymbol{B}}_5 \times \boldsymbol{E},
\end{align}
in the quantum regime.
As a result,
the regular Hall effect induces spin accumulation localized at the magnetic texture,
leading to the switching of magnetization via the spin torque.
In contrast to the conventional spin transfer torque driven by conduction current,
the spin torque noted here does not require a conduction current.
Therefore, although the Weyl electrons cannot be transmitted through a sharp magnetic texture,
as the valleys are shifted in momentum space in accordance with the magnetization,
this spin torque is still present and drives a motion of the magnetic texture.
This effect can also be regarded macroscopically as the electric driving of the localized charge $\rho_B$ shown above
attached to the magnetic texture \cite{Araki_2016_2,Araki_2018_2}.

In magnetic Weyl semimetals,
it is also proposed that an external magnetic field $\boldsymbol{B}$ under a gate voltage
induces a spin torque,
which is termed charge (voltage)-induced spin torque \cite{Nomura_2015,Kurebayashi_2016}.
This effect can be described in terms of the chiral magnetic effect:
when the Fermi level is lifted by $\delta \mu$ due to the gate voltage,
the magnetic field $\boldsymbol{B}$ induces the axial current
\begin{align}
  \boldsymbol{j}_5^{\mathrm{(C)}} = \boldsymbol{j}_+^{\mathrm{(C)}} - \boldsymbol{j}_-^{\mathrm{(C)}} = -\frac{e^2}{2\pi^2} \delta \mu \boldsymbol{B},
\end{align}
which is the axial current counterpart of the chiral magnetic effect
(see Figure \ref{fig:induced-current}(a)).
Therefore, if the gate voltage $\delta \mu$ is applied in a limited area,
the spin accumulation, corresponding to the axial current $\boldsymbol{j}_5^{\mathrm{(C)}}$,
enables one to switch the magnetization within this area,
without driving any electric current.

\section{Example: magnetic domain walls \label{sec:domain-wall}}
Based on the general theory about magnetic texture dynamics in Weyl semimetals,
let us focus on the effect of magnetic domain walls in this section,
as a typical example.
There have been several theoretical works on the electron dynamics and transport
in the presence of magnetic domain walls in magnetic Weyl semimetals.
It was seen both analytically and numerically that a magnetic domain wall in a Weyl semimetal
gives rise to a large domain-wall magnetoresistance,
due to the mismatch of the electron helicity beyond the domain wall \cite{Ominato_2017,Kobayashi_2018}.

One of the peculiar features of magnetic domain walls in Weyl semimetals
is the emergence of one-dimensional zero modes localized at the domain wall \cite{Araki_2016_2,Araki_2018_2}.
These zero modes can be regarded as the remnant of the surface Fermi arc of Weyl semimetal.
Macroscopically, this localized mode corresponds to the Landau states under the axial magnetic field from the domain wall texture.
Let us here see this correspondence by using a typical one-dimensional domain wall structure.

For a one-dimensional magnetic domain wall, several types of internal structure are possible.
If the domain wall is centered at $x=0$
and the magnetization in each region separated by the domain wall points to the direction parallel to the wall,
i.e. $\boldsymbol{M}(x \rightarrow \infty) = \pm M_0 \boldsymbol{e}_z$,
the internal structure of the domain wall can be formulated as
\begin{align}
  \boldsymbol{M}(x) = M_0 \left(\mathrm{sech}\frac{x}{w}\cos\alpha , \mathrm{sech}\frac{x}{w}\sin\alpha , \tanh \frac{x}{w} \right), \label{eq:domain-wall}
\end{align}
where the length scale $w$ corresponds to the thickness of the domain wall.
The internal structure is characterized by the angle $\alpha$:
the N\'{e}el domain wall, in which the magnetization is twisted in a coplanar manner (within $xz$-plane),
corresponds to $\alpha = 0,\pi$,
whereas the Bloch domain wall, in which the magnetization is twisted transverse to $x$-axis (within $yz$-plane),
corresponds to $\alpha = \pm \pi/2$.
In realistic magentic materials, the internal structure of the domain wall
is governed by the magnetic anisotropy and the Dzyaloshinskii--Moriya interaction
(under the broken inversion symmetry).
The head-to-head domain wall
$\boldsymbol{M}(x) = M_0 (\pm \tanh (x/w),0,\mathrm{sech}(x/w))$ can also be considered,
which we will not go into details in this article.

\begin{figure}[tbp]
  \includegraphics[width=\columnwidth]{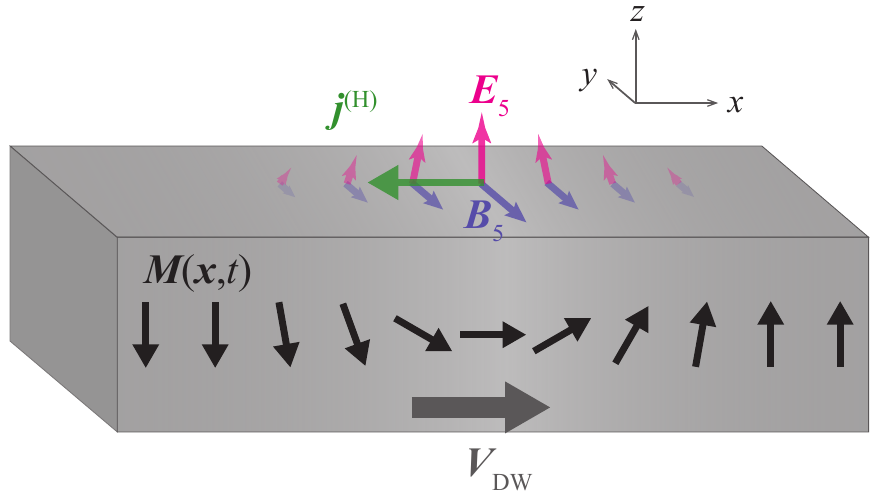}%
  \caption{\label{fig:DW-setup}
    Setup of the domain wall texture $\boldsymbol{M}(x,t)$ moving with velocity $V_{\mathrm{DW}}$, its corresponding axial electromagnetic fields $(\boldsymbol{E}_5,\boldsymbol{B}_5)$, and the induced regular Hall current $\boldsymbol{j}^{\mathrm{(H)}}$ that pumps the charge localized at the domain wall.}
\end{figure}

Under this domain wall, the Weyl electrons feel the axial magnetic field [Equation (\ref{eq:B5})]
\begin{align}
  \boldsymbol{B}_5(x) &= -\frac{J}{e v_{\mathrm{F}}} \boldsymbol{\nabla} \times \boldsymbol{M}(x) \label{eq:B5-DW}\\
  &= \frac{JM_0}{e v_{\mathrm{F}}w} \mathrm{sech}^2\frac{x}{w} \left( 0, 1, \sinh\frac{x}{w} \sin\alpha \right), \nonumber
\end{align}
which is localized around the domain wall at $x=0$ (see Figure \ref{fig:DW-setup}).
As seen in the previous section, this axial magnetic field leads to the Landau quantization
and gives rise to the locaized charge $\rho_B(x)$ around the domain wall.
If we assume that the Fermi level $\mu$ is in the quantum regime,
i.e. $\mu$ is between the zeroth and first Landau levels \textit{throughout the whole system},
only the zeroth Landau level contributes to the localized charge
and the induced charge density obeys Equation (\ref{eq:localized-charge}).
Thus the localized charge per unit area of the domain wall can be estimated as
\begin{align}
  q = \frac{e}{\pi^2} \frac{JM_0}{v_{\mathrm{F}}^2} \mu \ \text{(N\'{e}el)}, \quad 
  \frac{e}{2\pi} \frac{JM_0}{v_{\mathrm{F}}^2} \mu \ \text{(Bloch)},
\end{align}
for each type of the domain wall \cite{Araki_2018}.
This result was verified analytically for N\'{e}el domain wall using the Jackiw--Rebbi formalism \cite{Araki_2016_2},
and numerically for both types of domain walls using the lattice model \cite{Araki_2018_2},
which implies that the zeroth Landau level considered in this macroscopic description
really has a dominant contribution to the charging of domain walls.
It was seen in these literatures that the localized modes contributing to the charging
show the band structure similar to the Fermi arcs {on} the surface,
crossing the zero-energy plane by an open countour that bridges two Weyl points in momentum space.
The equilibrium current $\boldsymbol{j}^{\mathrm{(C)}}$ from the chiral pseudomagnetic effect [Equation (\ref{eq:j-chiral})]
was also explicitly seen by using the wave functions of the localized modes.

What will occur to this localized charge if the domain wall is moving?
In order to see the dynamical behavior,
here I assume that the domain wall is parallelly moving with velocity $V_{\mathrm{DW}}$ for simplicity,
which is reproduced by substituting the position $x$ in Equation (\ref{eq:domain-wall}) by $x-V_{\mathrm{DW}} t (\equiv x')$.
Such a domain wall motion can be driven by, for example,
applying a magnetic field externally.
The motion of the domain wall yields both the axial magnetic field,
given by substitution $x \rightarrow x'$ in Equation (\ref{eq:B5-DW}),
and the axial electric field
\begin{align}
  &\boldsymbol{E}_5(x') = \frac{J}{e v_{\mathrm{F}}} \partial_t \boldsymbol{M}(x') = -\frac{JV_{\mathrm{DW}}}{e v_{\mathrm{F}}} \partial_{x'} \boldsymbol{M}(x')\label{eq:E5-DW} \\
  & \quad = \frac{JM_0 V_{\mathrm{DW}}}{e v_{\mathrm{F}} w}\mathrm{sech}^2\frac{x'}{w} \left( \sinh\frac{x'}{w}\cos\alpha,\sinh\frac{x'}{w}\sin\alpha,-1 \right). \nonumber
\end{align}
Since $\boldsymbol{B}_5$ and $\boldsymbol{E}_5$ are perpendicular to one another,
they give rise to the regular Hall current $\boldsymbol{j}^{\mathrm{(H)}}$ localized at the domain wall
(see Figure \ref{fig:DW-setup}).
Among the regular Hall current, its $x$-component
\begin{align}
  j^{\mathrm{(H)}}_x(x') = -\frac{e JM_0 V_{\mathrm{DW}}\mu}{v_{\mathrm{F}}^2 w} \mathrm{sech}^2\frac{x'}{w} \sqrt{1+\sinh^2\frac{x'}{w}\sin^2\alpha}
\end{align}
contributes to the pumping of the localized charge,
satisfying the charge conservation $\partial_t \rho_B + \partial_x j^{\mathrm{(H)}}_x =0$.
Therefore, we can see from the macroscopic axial electromagnetic field picture
that a magnetic domain wall in the Weyl semimetal bears
a certain amount of localized charge depending on the internal structure of the domain wall,
and that the localized charge is carried along with the motion of the domain wall,
which can be regarded as the regular Hall current driven by the axial electromagnetic fields.

{The localized modes discussed here can also be understood
in connection with the Jackiw--Rebbi mode,
which arises as the zero-energy solitonic solution of the Dirac equation
localized at a domain wall in the Dirac mass term \cite{Jackiw_1976}.
Let us consider the case of a N\'{e}el domain wall ($\alpha = 0$ in Equation (\ref{eq:domain-wall})).
By taking the transverse momentum components $k_y$ to zero and $k_z$ to a fixed value,
the Weyl Hamiltonian coupled with the magnetic texture reduces to a one-dimensional Dirac Hamiltonian along $x$,
\begin{align}
  H_{\eta,k_z} &= \left[-i v_{\mathrm{F}} \eta \partial_x + JM_0 \mathrm{sech}\frac{x}{w}\right] \sigma_x + m_{\eta,k_z}(x) \sigma_z,
\end{align}
with the $x$-dependent Dirac mass term $m_{\eta,k_z}(x) = \eta v_{\mathrm{F}} k_z  + JM_0 \tanh\frac{x}{w}$.
By the U(1) gauge transformation with $U(x) = \exp \left[ -i\eta \frac{JM_0}{v_{\mathrm{F}}} \int dx \ \mathrm{sech}\frac{x}{w} \right]$,
the Hamiltonian can be further reduced as $H_{\eta,k_z} = -i v_{\mathrm{F}} \eta \partial_x\sigma_x + m_{\eta,k_z}(x) \sigma_z$,
to which one can apply Jackiw--Rebbi's discussion.
The localized zero mode arises when $m_{\eta,k_z}(x)$ changes its sign at $x=0$,
which occurs if $|k_z| < JM_0/v_{\mathrm{F}}$,
corresponding to the condition for the surface Fermi arc.
While the Jackiw--Rebbi mode interpretation is successful for the N\'{e}el domain wall,
it is not applicable to the Bloch domain wall,
since the Hamiltonian cannot exactly be reduced to the Jackiw--Rebbi form by the gauge transformation
due to the $z$-component in $\boldsymbol{B}_5$ (see Equation (\ref{eq:B5-DW})).
Thus we need correction by the pseudomagnetic fields to Jackiw--Rebbi's discussion,
when we investigate spin-momentum-locked electrons under magnetic domain walls.
}

It is suggested that the localized charge on the domain wall can serve as a ``knob'' for the manipulation of the domain wall motion by an external electric field;
this electric manipulation was numerically simulated by solving the Landau--Lifshitz--Gilbert equation on the lattice model,
using the spin torques mentioned in Section \ref{sec:torque} \cite{Kurebayashi_2019}.

\section{Magnetic textures in various Weyl semimetals \label{sec:realistic}}
So far we have taken a fully spin-momentum locked Weyl semimetal model,
namely the minimal model containing $\boldsymbol{\sigma} \cdot \boldsymbol{k}$ term,
to get an intuitive understanding on the interplay between spin-momentum locking of the electrons
and real-space magnetic textures.
In realistic Weyl semimetal materials, on the other hand,
the spin-momentum locking structure around the Weyl points is material-dependent,
since the band crossing comes from spin-orbit coupling characteristic to each material.
As I have mentioned in the first Section,
significant progress in the material-based first-principle calculations
and the experimental measurement techniques
has enabled us to observe several compounds as candidates for Weyl semimetals.
In this section,
I review the current status of theoretical and experimental works on realizing ferromagnetic/antiferromagnetic Weyl semimetals,
and see how magnetic textures and the electronic properties are related in these materials.

\subsection{Topological insulator multilayers}
From the early days of the theoretical works on Weyl semimetals,
it has been proposed that the ferromagnetic Weyl semimetal phase can be realized in magnetic multilayers
composed of thin films of magnetically-doped topological insulator (TI) and normal insulator (NI) \cite{Burkov_2011,Burkov_2011_2}.
When a 3D TI is magnetically doped,
its surface Dirac cone becomes gapped out due to the breaking of time-reversal symmetry by the ferromagnetic order,
leading to the quantum anomalous Hall effect \cite{Haldane_1988,Yu_2010,Chang_2013}.
By stacking these TI films with NI films in between,
the top and bottom surface states of the TI films are moderately coupled due to intralayer and interlayer tunneling,
leading to the band dispersion perpendicular to the layers.
Depending on the ratio between the intralayer and interlayer tunneling amplitudes,
the system can exhibit 3D TI and NI phases.
Between these phases emerges the Weyl semimetal phase,
with closing of the band gap at the Weyl points \cite{Murakami_2007}.

The Hamiltonian around the Weyl points can be symbolically written as
\begin{align}
  H(\boldsymbol{k}) = v_{\mathrm{F}} (\boldsymbol{e}_z \times \boldsymbol{\sigma}) \cdot \boldsymbol{k} +m(k_z) \sigma_z,
\end{align}
where $\sigma_{x,y,z}$ are the spin Pauli matrices mixed with orbital degrees of freedom.
The function $m(k_z)$, governed by the tunneling amplitudes and the exchnge energy by the magnetic dopants,
reaches zero at two $k_z$ points, corresponding to the Weyl points.
In this Hamiltonian,
the in-plane spin-momentum locking is inherited from that of the TI surface states and hence is Rashba-like \cite{Zhang_2009,Liu_2010}.
Therefore, an in-plane magnetization leads to an effective vector potential perpendicular to the magnetization direction.
On the other hand, the out-of-plane magnetization gives rise to shift of the Weyl points along $k_z$-axis via the $m(k_z)$-term,
yielding an out-of-plane effective vector potential.
Since all three components of spin are coupled to momentum,
the magnetic textures formed by the magnetic dopants in this system
lead to the charge and current profile similar to those listed up in Section \ref{sec:Weyl-case},
except for the in-plane spin texture here being twisted by 90 degrees from that under the $(\boldsymbol{k} \cdot \boldsymbol{\sigma})$-Hamiltonian,
due to the Rashba-like spin-momentum locking.
(Since the spin Pauli matrices here are mixed with orbital degrees of freedom,
the effective vector potentials mentioned here should be viewed as the mixture of the normal and axial vector potentials.)

Similarly to the TI multilayers,
it is proposed that a magnetically doped bulk 3D TI can also exhibit the Weyl semimetal phase
provided that the doping ratio is properly tuned:
the Weyl phase again emerges as the intermediate phase between the TI and NI phases \cite{Kurebayashi_2014}.
From the mean-field theory,
it was shown that about 10 percent doping of Cr in $\mathrm{Bi}_2 (\mathrm{Se}_x \mathrm{Te}_{1-x})_3$ can drive the system from the TI phase
into the Weyl phase.
Since it is based on the band inversion by spin-orbit coupling in TI,
the spin-momentum locking structure in this system is similar to that in the above multilayer system,
where the in-plane part is given by the Rashba-like form.

\subsection{Layered kagome ferromagnet: $\mathrm{Co_3 Sn_2 S_2}$}
Recent studies suggested that a cobalt-based shandite material $\mathrm{Co_3 Sn_2 S_2}$
works well as a Weyl semimetal with the ferromagnetic order \cite{Liu_2018,Xu_2018,Wang_2018}.
The ferromagnetic order consists of the magnetic moments in Co atoms arranged in kagome-lattice layers,
with the out-of-plane magnetic moment $\sim 0.3 \mu_B$ for each Co atom \cite{Kassem_2016}.
The Weyl points due to the breaking of time-reversal symmetry reside just $60 \mathrm{meV}$ beyond the Fermi energy,
with no other bands crossing the Fermi level \cite{Liu_2018}.
Thus $\mathrm{Co_3 Sn_2 S_2}$ may serve as a good platform for realizing the anomalous transport characteristic to Weyl semimetals:
for instance, first-principle calculations and transport measurements show
both a large anoumalous Hall conductivity up to $\sim 10^3 \Omega^{-1} \mathrm{cm}^{-1}$
and a large anomalous Hall angle (the ratio of the anomalous Hall conductivity to the longitudinal conductivity) up to $\sim 20$,
which are not simultaneously reached in conventional magnetic materials \cite{Liu_2018}.
The Weyl band structure and the emergent Fermi arcs on surface were also observed
by angle-resolved photoemission spectroscopy (ARPES) measurements \cite{Wang_2018}.

Identification of magnetic textures in kagome magnets
is also an important question from the viewpoint of magnetism,
as the kagome lattice can host various kinds of unconventional spin correlations,
such as the spin frustration and Dzyaloshinskii--Moriya interaction.
From the measurement of magnetization and ac susceptibility by superconducting quantum interference device (SQUID),
an anomalous magnetic phase (A-phase) distinct from the ferromagnetic phase was observed in $\mathrm{Co_3 Sn_2 S_2}$,
which appears just below the Curie temperature $T_C = 173 \mathrm{K}$ at low magnetic field \cite{Kassem_2017}.
It was suggested that this phase may host {nonuniform spin textures such as stripe domains, magnetic bubbles, or biskyrmions.}
Further observation of magnetic textures in the A-phase,
by neutron scattering, magnetic force microscopy, etc.,
is now pending.

Due to its layered structure,
spin-orbit coupling in $\mathrm{Co_3 Sn_2 S_2}$ is higly anisotropic,
which may result in an anisotropy in the spin-momentum locking structure around the Weyl points.
It was suggested by the first-principles calculations
that a rotation of the magnetization from the out-of-plane direction to the in-plane direction eventually modulates the arrangement of the Weyl nodes,
with their complicated trajectories in the Brillouin zone \cite{Ghimire_2019}.
It was also theoretically confirmed
that a simplified two-orbital tight-binding model on kagome lattice
with Kane--Mele-type spin-orbit interaction
well reproduces the Weyl-node structure observed in the first-principles calculations \cite{Ozawa_2019};
their agreement implies that the out-of-plane spin component dominantly participates in spin-orbit coupling in $\mathrm{Co_3 Sn_2 S_2}$,
in the vicinity of the Fermi level.
As a result, we can roughly expect that
the out-of-plane component of the magnetization shifts the Weyl nodes
and contributes to the pseudo-gauge field,
as we have seen in the spin-momentum-locked model Hamiltonian,
while the in-plane component couples to the electron spin conventionally as in normal metals.
The effect of the experimentally-suggested magnetic textures mentioned above
on the electronic structure and transport
is left as an important question for future spintronics application of this material.

\subsection{Noncollinear antiferromagnet: $\mathrm{Mn_3 Sn}$}
While we have so far focused on the Weyl semimetals
with time-reversal symmetry broken by ferromagnetism,
we may also consider antiferromagnets with Weyl-node structure.
$\mathrm{Mn_3 Sn}$ and $\mathrm{Mn_3 Ge}$ are known to exhibit both antiferromagnetism and the Weyl nodes,
which was first proposed by first-principles calculations \cite{Chen_2014}
and later manifestly observed by transport and ARPES measurements \cite{Nakatsuji_2015,Nakatsuji_2016,Nayak_2016,Ikhlas_2017,Kuroda_2017}.
Mn atoms are responsible for the antiferromagnetic order in these materials,
arranged in kagome-lattice layers.
The magnetic moments on these kagome sites form
a $120$-degree noncollinear antiferromagnetic order,
with the N\'{e}el temperature up to $\sim 430$K in $\mathrm{Mn_3 Sn}$ \cite{Kren_1975,Nagamiya_1982,Tomiyoshi_1982,Brown_1990}.
It is understood that this noncollinear antiferromagnetism arises from the combination of the antiferromagnetic excchange coupling and the Dzyaloshinski--Moriya interaction.
Although their net magnetization is vanishingly small
($\sim 0.002 \mu_B$ for each Mn atom) \cite{Nagamiya_1982,Tomiyoshi_1982}
and the scalar spin chirality is zero,
it was numerically and experimentally observed that these materials show large anomalous Hall conductivity \cite{Nakatsuji_2015,Nakatsuji_2016,Nayak_2016},
which can be regarded as an effect from the strong Berry flux
in momentum space around the Weyl points.
A tight-binding model calculation on kagome layers confirmed that
the $k$-space Berry curvature indeed reproduces
this large anomalous Hall conductivity around the Fermi level \cite{Ito_2017}.
{Although there exist trivial metallic bands in addition to the Weyl cones at the Fermi level
so that they cannot be regarded as ``semimetals'',
the Weyl points thus contribute strongly to the anomalous transport properties.}

One can consider a spatial modulation of this 120-degree pattern
at macroscopic length scales much longer than the lattice constant.
Since there is no unified order parameter such as magnetization in ferromagnets and N\'{e}el vector (staggered magnetization) in collinear antiferromagnets,
one should pick up some other vectorial quantity that characterizes the macroscopic magnetic textures.
One powerful quantity that we can rely on is the cluster octupole moment,
which was derived by the multipole expansion in each cluster \cite{Suzuki_2017}.
It has the same symmetry as the magnetic dipole moment,
and gives rise to the anomalous Hall effect in the same manner with dipole.
Domain structure identified by the cluster octupole moment was
experimentally distinguished by measuring the magneto-optical Kerr effect (MOKE) \cite{Higo_2018}.

Since the spin-momentum locking structure is complex due to the noncollinear spin texture,
it seems difficult to apply the pseudo-gauge field picture to treat the effect of macroscopic magnetic textures.
One way to resolve this problem theoretically
is to list up the possible terms allowed by the symmetry of the system.
In Reference \cite{Liu_2017}, the structure of domain wall in the kagome antiferromagnet was
parameterized by the in-plane twisting angle of the noncollinear order,
and the current-induced domain-wall dynamics was qualitatively discussed
by using the collective force on the domain wall allowed by the symmetry.
The orbital magnetization may be helpful to understand the qualitative effect of magnetic textures on the shift of Weyl points,
since they are closely related via the intrinsic anomalous Hall conductivity \cite{Streda_1982,Koshino_2016}.
It was seen by the tight-binding model calculation that
the direction and magnitude of the orbital magnetization significantly depend
on the tilting angles of the noncollinear spins \cite{Ito_2017}.
Thus one can assume that a twist in the noncollinear spin texture leads to shift of Weyl points,
giving rise to the pseudo-gauge field texture.

\section{Conclusion} \label{sec:conclusion}
In this article, we have reviewed the current understanding of
the interplay between magnetic textures and electronic properties in magnetic Weyl semimetals.
Since the magnetization contributes to the shift of Weyl points in momentum space,
the effect of magnetic textures on the electrons can be effectively treated
as axial gauge fields,
with which one can macroscopically treat the electronic structure and transport
in a similar manner with
those under the realistic electromagnetic fields.
In contrast to the spin gauge fields $\boldsymbol{\mathcal{A}}(\boldsymbol{r})$ in normal metals,
which depend on spatial gradients in the magnetization pattern $\boldsymbol{\nabla} \boldsymbol{M}(\boldsymbol{r})$ (Section \ref{sec:spin-texture}),
the axial gauge fields $\boldsymbol{A}_5(\boldsymbol{r})$ for the Weyl electrons is directly related to the local magnetization $\boldsymbol{M}(\boldsymbol{r})$ due to spin-momentum locking (Section \ref{sec:Weyl-case}).
From this difference, we have seen that the electronic structure in Weyl semimetals gets largely altered by magnetic textures;
even a simple one-dimensional domain wall leads to the localization and pumping of electric charge,
as we have seen in Section \ref{sec:domain-wall}.

We have mainly focused on the effect of long-range and low-frequency magnetic textures introduced by hand in this article.
On the other hand, magnetically ordered materials generally host spin-wave excitations, namely magnons,
which arise as collective dynamics in the precession of spins interacting with each other.
It is well known that magnons carry spin current \cite{Kajiwara_2010},
and hence their properties are being intensely studied
to make use of them as carriers in spintronics devices,
generically termed as magnonics \cite{Kruglyak_2010,Chumak_2015}.
Since magnons can couple to spin-momentum-locked Weyl electrons
in magnetic Weyl semimetals,
there have been several theoretical proposals
about magnon-induced anomalous responses \cite{Liu_2013,Hutasoit_2014,Araki_2016}.
Interplay between electrons and magnons in realistic Weyl semimetals
needs to be questioned in future studies \cite{Shvetsov_2019}.





\textbf{Acknowledgments.}
This work is supported by JSPS KAKENHI Grant Number JP17K14316.
The author acknowledges K. Nomura and A. Yoshida for the related collaborations.
The author thanks D. Kurebayashi for the fruitful discussions.

\textbf{Conflict of interest.}
The author declares no conflict of interest.

\end{document}